\documentclass[]{pasj01}
\draft
\begin{document} 
\Received{}
\Accepted{}
\title{Characterization of diffuse X-ray emission from IGR~J17448$-$3232: an implication of a line of sight merging activity}

\author{
Shoko \textsc{Watanabe}\altaffilmark{1}, Shigeo \textsc{Yamauchi}\altaffilmark{1$\ast$}, Kumiko K. \textsc{Nobukawa}\altaffilmark{1}, 
and 
Hiroki \textsc{Akamatsu}\altaffilmark{2}
}
\altaffiltext{1}{Faculty of Science, Nara Women's University, Kitauoyanishimachi, Nara 630-8506}
\email{yamauchi@cc.nara-wu.ac.jp}
\altaffiltext{2}{SRON Netherlands Institute for Space Research, Sorbonnelaan 2, 3584 CA Utrecht, The Netherlands}
\maketitle

\begin{abstract}
Results of the spectral analysis for the galaxy cluster IGR J17448$-$3232 are presented. 
The intracluster medium (ICM) in the central region ($r<$300$''$, 320 kpc) has a high electron temperature plasma of $kT_{\rm e}$$\sim$13--15 keV and  
an ionization temperature estimated from an intensity ratio of Fe\emissiontype{XXVI} Ly$\alpha$/Fe\emissiontype{XXV} He$\alpha$ lines is 
lower than the electron temperature, which suggests that the ICM is in the non-ionization equilibrium (NEI) state. 
The spectrum in the central region can be also fitted with a two-component model:  
a two-temperature plasma model in a collisional ionization equilibrium (CIE) with temperatures of 7.9 keV and $>$34 keV 
or a CIE$+$power law model with a temperature of 9.4 keV and a photon index of 1.1. 
The two component models can represent the intensity ratio of Fe\emissiontype{XXVI} Ly$\alpha$/Fe\emissiontype{XXV} He$\alpha$ lines.
On the other hand, the spectrum in the outer region ($r>$300$''$) can be explained by a single CIE plasma model with a temperature of 5--8 keV. 
Based on the spectral feature and its circular structure, we propose 
that the NEI plasma was produced by merging along the line-of-sight direction.
\end{abstract}

\section{Introduction}

Galaxy clusters, consisting of galaxies, thin hot plasma, and dark matter, are the largest virialized objects in the Universe. 
The galaxy clusters have grown primarily via merging surrounding subclusters and accretion flow from large-scale structures. 
Some evidence of the past merging is found in their spatial and temperature structures: 
the shock wave produced by the merging process heats intracluster medium (ICM), and hence makes a very hot plasma
(e.g., $kT_{\rm e}\sim$15 keV for 1E 0657$-$56: \cite{Markevitch2002}), while the ICM exhibits a non-circular structure.
From dynamical (equilibration) time scale ($\sim$10$^9$ yr) and their low-density ($n_{\rm e}\sim10^{-2}$--$10^{-3}$ cm$^{-3}$) nature in the central region, 
the ICM would be in a collisional ionization equilibrium (CIE). 
In fact, the CIE model well represents spectra of the ICM. 

Theoretical researches predict that the ICM is in a non-equilibrium ionization (NEI) state after a merging process 
(e.g., \cite{Takizawa1999,Akahori2010}). 
\citet{Fujita2008} examined an ionization temperature using 
an intensity ratio of Fe\emissiontype{XXVI} Ly$\alpha$/Fe\emissiontype{XXV} He$\alpha$ lines of the Ophiuchus galaxy cluster, 
one of the hottest known galaxy clusters, 
but found that the ICM has reached the ionization equilibrium state. 
Recently, \citet{Inoue2016} reported that an ionization temperature derived from 
the intensity ratio is below an electron temperature in the hottest part of A754, 
which suggests that the ICM is in the NEI state. 
There are also some arguments that no significant evidence of the NEI plasma is found even in merging clusters (e.g., \cite{Russell2012}).
Thus, it is now debatable whether galaxy clusters contain NEI plasmas or not.

IGR J17448$-$3232 is a source located near to the Galactic plane [($l$, $b$)$\sim$(\timeform{356.D8}, \timeform{-1.D7})], 
discovered by INTEGRAL (Bird et al. 2007, 2010). 
Chandra observed the field including IGR J17448$-$3232 and found an extended source, CXOU J174453.4$-$323254, 
which is a counterpart of IGR J17448$-$3232 \citep{Tomsick2009}.
Follow up observations with XMM-Newton revealed that IGR J17448$-$3232 is a galaxy cluster hidden behind the Galactic bulge \citep{Barriere2015}.
The redshift was estimated to be $z$=0.055$\pm$0.001 from the X-ray spectroscopy.
The X-ray emission exhibits roughly a circular structure and the radial profile is well represented by an isothermal $\beta$ model 
with a scale radius of \timeform{3'.27} and $\beta$ of 0.56 \citep{Barriere2015}.
The temperature is quite high ($kT_{\rm e}>$10 keV) in the inner parts, 
and gradually decrease to $kT_{\rm e}$=4 keV at the 10$'$--13$'$ annulus region as is apart from the center \citep{Barriere2015}.

The existence of the high temperature plasma may suggest that IGR J17448$-$3232 is in the post merger phase.
As mentioned above, IGR J17448$-$3232 has several striking features including a large core radius, 
and spherical symmetric X-ray distribution, which is non-trivial in the merging clusters. 
To understand the orgin of these features, further X-ray analysis is required.
Thus, we analyzed the XMM-Newton data and found a sign that an ionization temperature derived from an intensity ratio 
of Fe\emissiontype{XXVI} Ly$\alpha$/Fe\emissiontype{XXV} He$\alpha$ lines does not match with the electron temperature.
In this paper, we report results of spectral analysis. 
We utilize cosmological parameters of $\Omega_{\rm m}$=0.3, $\Omega_{\Lambda}$=0.7, and H$_0$=70 km s$^{-1}$ Mpc$^{-1}$. 
Adopting a redshift of $z$=0.055 \citep{Barriere2015}, 1$'$=64 kpc. 
The quoted errors are at the 90\% confidence level unless otherwise mentioned.

\section{Observation and Data Reduction}

Observation of the galaxy cluster IGR J17448$-$3232 
was carried out with the European Photon Imaging Camera (EPIC) on board XMM-Newton (OBSID 0672260101).
The EPIC system is composed of two different detectors; 
two MOS cameras (MOS 1 and MOS 2: \cite{Turner2001}) and one pn camera
\citep{Struder2001}. 
The MOS and pn cameras were operated in the full-frame mode. 
The spatial resolution of the XMM-Newton X-ray telescopes is $\sim$15$''$.

Data reduction and analysis were made using the ESAS (Extended Source Analysis Software) package \citep{Snowden2008} version 15.0.0 
and HEASOFT version 6.24. 
We excluded the data including background flare events. 
The resultant exposure times are 16.8, 25.2, and 26.5 ks for pn, MOS 1, and MOS 2, respectively. 
In the present observation, MOS 1 CCD6 and MOS 2 CCD5 are out-of-function. 
Within 300$''$ region, pn, MOS 1 and MOS 2 are available, but 
for the analysis of the spectra in the outer region ($r>$300$''$), only the pn data are used (see figure 1).

\section{Analysis and Results}

\subsection{Spatical distribuion}

\begin{figure}
  \begin{center}
\includegraphics[width=8cm]{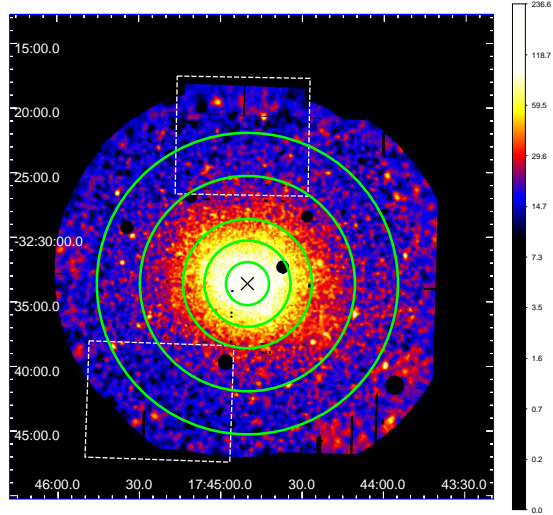} 
  \end{center}
  \caption{
EPIC image of IGR J17448$-$3232 in the 0.5--10.0 keV energy band (color scale). 
The data of pn, MOS 1, and MOS 2 were co-added. 
The background subtraction and the exposure correction are made.
The coordinates are J2000.0.   
The color bar shows intensity levels in the arbitrary unit.
Contributions of point sources are excluded. 
The black cross shows the peak position of the X-ray emission. 
MOS 1 CCD6 and MOS 2 CCD5 were out-of-function in the present observation. 
The areas of the CCD chips are shown by white dashed lines.
The regions from which the source spectra are extracted are shown by the green lines.
}\label{fig:img}
\end{figure}

Figure 1 is an X-ray image of IGR J17448$-$3232 in the 0.5--10.0 keV energy band. 
Contributions of point sources which have been reported in \citet{Barriere2015} are excluded.
In order to maximize photon statistics, the data of MOS 1, MOS 2, and pn are co-added.
It clearly shows a single peak structure with no substructures. 
The peak position is estimated to be ($\alpha$, $\delta$)$_{\rm J2000.0}$=(\timeform{266.D20888}, \timeform{-32.D56344}). 

We made a radial profile and confirmed that it is well represented by an isothermal $\beta$ model with 
a core radius, $r_{\rm c}$, of \timeform{3.'0}$\pm$\timeform{0.'5} and $\beta$ of 0.55$\pm$0.09.  
The resultant $\beta$ is consistent with those of relaxed clusters, 
on the other hand, $r_{\rm c}$ shows somewhat higher value than those of relaxed clusters (e.g., \cite{Akahori2005}).

\subsection{Spectral analysis}

\subsubsection{Estimation of the sky background}

Since IGR J17448$-$3232 is located near to the Galactic plane, 
we take account of the Galactic diffuse X-ray emission (GDXE) together with the Cosmic X-ray background (CXB) as a sky background. 
Thus, we at first estimate the sky background spectrum using data 
obtained on ($l$, $b$)=(\timeform{356.D4}, \timeform{-1.D5}) 
(the separation angle from IGR J17448$-$3232 of $\sim$\timeform{0.D4})
with the X-ray Imaging Spectrometer (XIS:  \cite{Koyama2007}) onboard Suzaku 
\citep{Mitsuda2007} (Obs. ID 505084010). 

\citet{Uchiyama2013} reported that the GDXE spectra are well represented 
by a model consisting of a foreground emission (FE), low- and high-temperature plasma components (LP and HP, respectively), 
and a reflection component (RC).
The FE is composed of two components, 0.09 keV and 0.59 keV plasmas.
The RC is composed of an absorbed power-law function (PL) and K$\alpha$ (6.400 keV) and K$\beta$ (7.058 keV) lines from neutral Fe. 
The sky background model is expressed as 
\begin{eqnarray}
\{{\rm TP}_{\rm FE}(kT_{\rm e}=0.09\ {\rm keV})+{\rm TP}_{\rm FE}(kT_{\rm e}=0.59\ {\rm keV})\}\times {\rm ABS}_1\nonumber \\
+ \{{\rm TP}_{\rm LP}+{\rm TP}_{\rm HP}+{\rm RC}\} \times {\rm ABS}_2+{\rm CXB} \times {\rm ABS}_3\hspace{1cm} 
\end{eqnarray}
where TP and ABS show a thermal plasma model ({\tt vapec} in XSPEC) and photoelectric absorption 
({\tt phabs} in XSPEC), respectively.
The K$\beta$ line intensity was fixed to be 0.125 $\times$ K$\alpha$ line intensity 
and the equivalent width of K$\alpha$ line was fixed to be 457 eV \citep{Uchiyama2013}.
The metal abundances, electron temperatures, and the $N_{\rm H}$ value for FE
were fixed to those in \citet{Uchiyama2013} and 
the spectral parameters of the CXB were fixed to the values in \citet{Kushino2002}. 
The $N_{\rm H}$ value of ABS$_3$ is assumed to be twice of that of ABS$_2$.
The abundance tables were taken from \citet{Anders1989}.
The cross sections of photoelectric absorption were taken from \citet{bcmc1992}. 

The best-fit spectral parameters are listed in table 1, while the best-fit model is shown in figure 2.
This model gave a reduced $\chi^2$ ($\chi^2$/d.o.f.) value of 296/214=1.38.
There are possible systematic errors, such as an energy scale, atomic data of Fe L-lines, and small calibration errors near the Si\emissiontype{I} K-edge energy. 
Taking account of these systematic errors, we use this model as a reasonable approximation. 

\begin{figure}
  \begin{center}
\includegraphics[width=8cm]{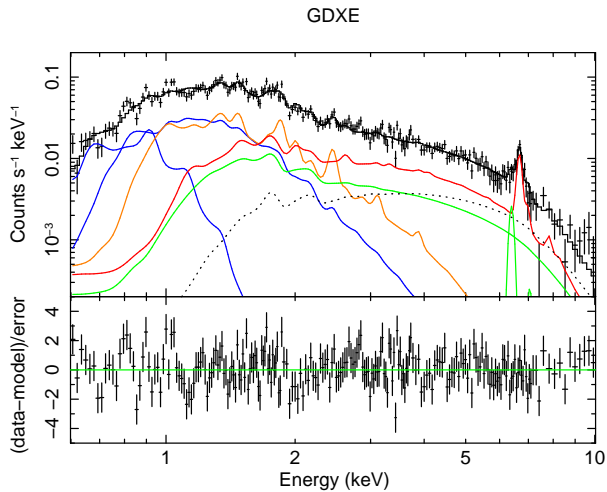} 
  \end{center}
  \caption{
Suzaku XIS spectrum of the nearby GDXE (upper panel) and residuals from the best-fit model (lower panel).
Errors of the data points are at the 1 $\sigma$ level.
The best-fit model is plotted by the histogram. The blue, orange, red, green lines show the FE, LP, HP, and RC, respectively, while the black dotted line shows the CXB
(see text and table 1).
}\label{fig:img}
\end{figure}

%
\begin{table}
\caption{The best-fit parameters of spectral analysis for the sky background, GDXE and CXB.}
\begin{center}
\begin{tabular}{llc} \hline  
Component & Parameter &Value\\ 
\hline 
Abs$_{\rm 1}$ & $N_{\rm H, FE}$ (cm$^{-2}$) & 5.6$\times10^{21}$ (fixed)\\
FE & $kT_{\rm e}$ (keV) &  0.09 (fixed)  \\
 & Normalization$^{\ast}$& (2.0$^{+0.3}_{-0.4}$)$\times$10$^{-2}$  \\ 
 & $kT_{\rm e}$ (keV) &  0.59 (fixed)  \\
 & Normalization$^{\ast}$& (2.9$^{+1.1}_{-1.4}$)$\times10^{-5}$  \\ 
  & Abundance $^{\dag}$  (solar)& 0.05 (fixed) \\
Abs$_{\rm 2}$ & $N_{\rm H, GDXE}$ (cm$^{-2}$) &  (1.3$^{+0.3}_{-0.2}$)$\times10^{22}$   \\
LP & $kT_{\rm e}$ (keV) & 0.88$^{+0.08}_{-0.11}$  \\
& Normalization$^{\ast}$& (1.2$\pm$0.3)$\times$10$^{-5}$  \\
& Abundance$^{\dag}$ (solar) & 1 (fixed) \\
HP & $kT_{\rm e}$ (keV) & 4.9$\pm$0.9 \\
& Normalization$^{\ast}$& (7.3$\pm$1.6)$\times10^{-6}$  \\
& Abundance$^{\dag}$ (solar) & =LP \\
RC& $\Gamma$ & 2.13 (fixed)  \\
& $I_{\rm 6.4 keV}^{\ddag}$ & (1.6$\pm$0.6)$\times$10$^{-8}$\\ 
& Equivalent width (eV) & 457 (fixed) \\ 
Abs$_{\rm 3}$ & $N_{\rm H, CXB}$ (cm$^{-2}$) & =2$\times$$N_{\rm H, GCXE}$ \\
CXB& $\Gamma$ &1.412 (fixed) \\
& Normalization$^{\S}$ &8.17$\times$10$^{-7}$ (fixed) \\ \hline
& $\chi^2$/d.o.f. & 296/215    \\
\hline \\
\end{tabular}
\end{center}
\vspace{-12pt}
$^{\ast}$ Defined as 10$^{-14}$$\times$$\int n_{\rm H} n_{\rm e} dV$ / (4$\pi D^2$ $\Omega$) (cm$^{-5}$ arcmin$^{-2}$),
where $n_{\rm H}$, $n_{\rm e}$, $D$, and $\Omega$ are hydrogen density (cm$^{-3}$), electron density (cm$^{-3}$), 
distance (cm), and solid angle (arcmin$^{-2}$), respectively. \\
$^{\dag}$ Relative to the solar value \citep{Anders1989}.\\
$^{\ddag}$ The unit is photons s$^{-1}$ cm$^{-2}$ arcmin$^{-2}$.\\
$^{\S}$ The unit is photons s$^{-1}$ cm$^{-2}$ keV$^{-1}$ arcmin$^{-2}$ at 1 keV.\\
\end{table}

\subsubsection{Spectrum of IGR J17448$-$3232}

\begin{figure*}
  \begin{center}
\includegraphics[width=8cm]{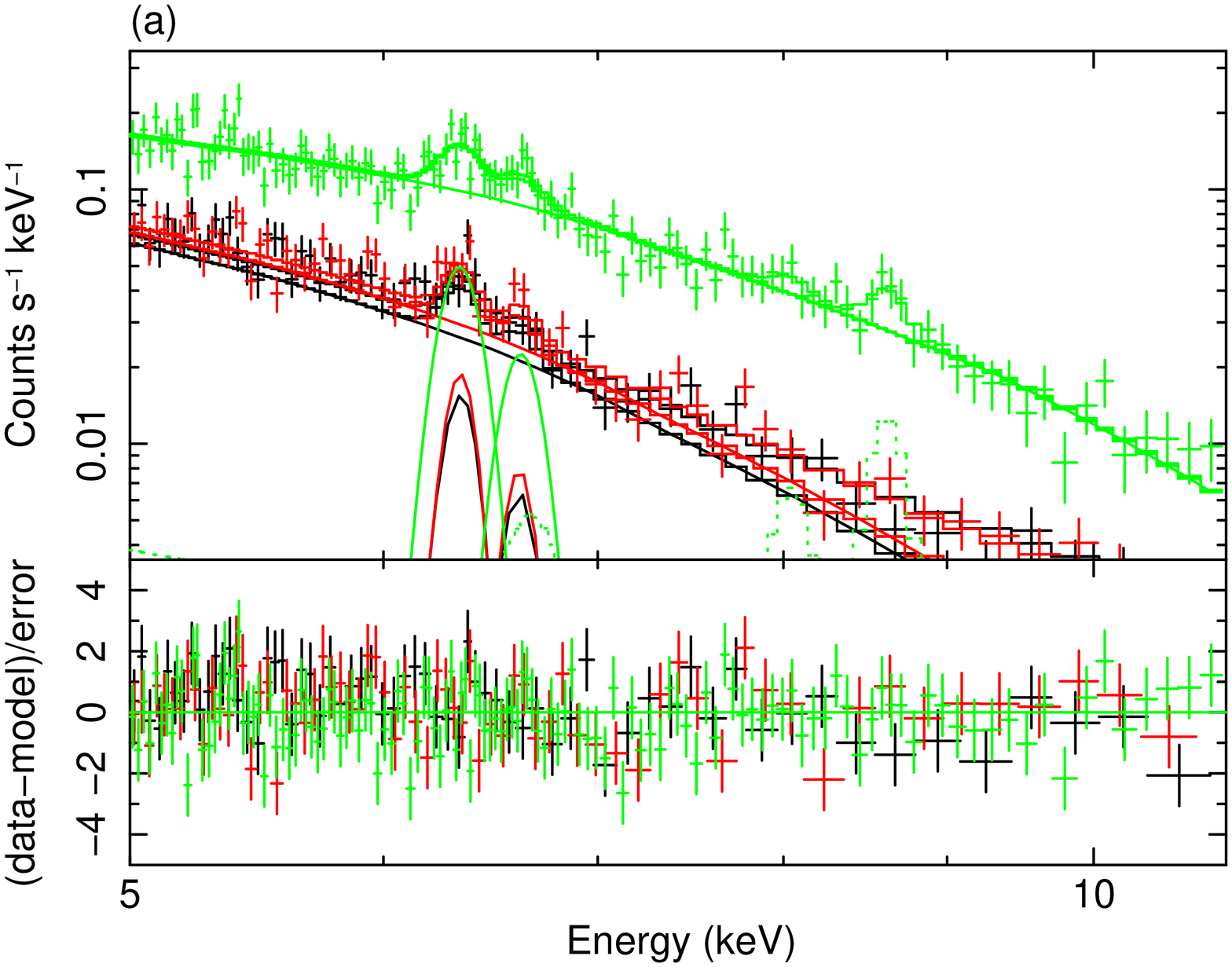} 
\includegraphics[width=8cm]{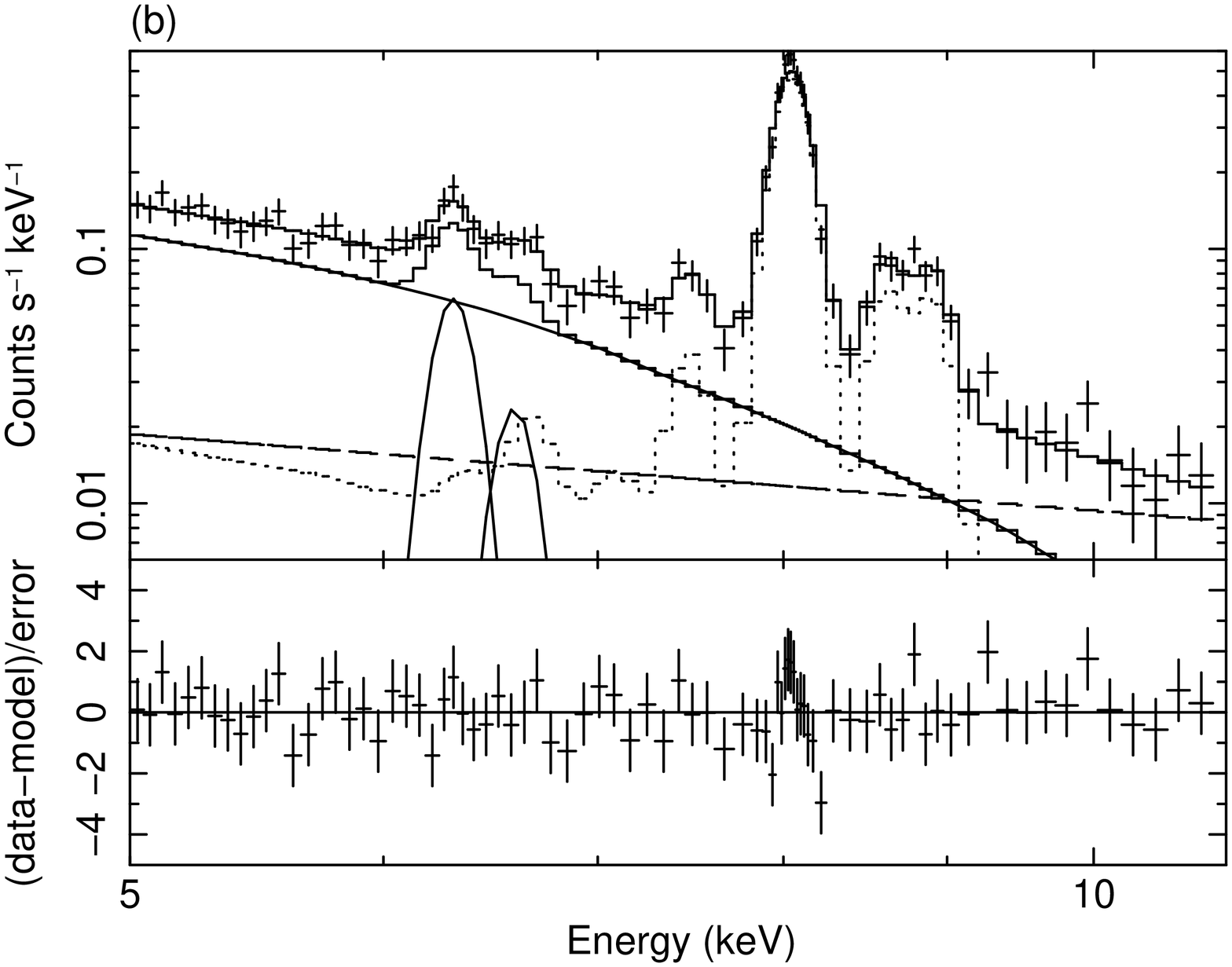} 
  \end{center}
  \caption{
(a)
X-ray spectrum of IGR J17448$-$3232 of 100$''$--200$''$  
(upper panel) and residuals from the best-fit model (lower panel). 
Black, red, and green colors show MOS 1, MOS 2, and pn, respectively. 
Errors of the data points are at the 1 $\sigma$ level. 
The solid and dotted lines show emission from IGR J17448$-$3232 (bremsstrahlung and gaussians) and sky background and instrumental lines, respectively. 
Although the model fitting was carried out for 1.2--11 keV band spectra, the 5--11 keV band data are displayed for the brevity. 
(b) Same as (a), but the pn spectrum of 300$''$--500$''$. 
The solid, dotted, and dashed lines show emission from IGR J17448$-$3232, 
sky background and instrumental lines, and a soft proton component, respectively.
}\label{fig:img}
\end{figure*}

%
\begin{table*}[t]
\caption{The best-fit parameters of a spectral analysis for IGR J17448$-$3232: bremsstrahlung$+$emission lines .}
\begin{center}
\begin{tabular}{lccccc} \hline  
Parameter &\multicolumn{5}{c}{Value}\\ 
&  0$''$--100$''$ & 100$''$--200$''$ & 200$''$--300$''$ &300$''$--500$''$ & 500$''$--700$''$  \\
\hline 
$N_{\rm H}$ ($\times10^{22}$ cm$^{-2}$) & 
2.06$^{+0.08}_{-0.05}$ 	& 2.06$\pm$0.04 		& 1.93$^{+0.07}_{-0.08}$ 	& 2.1$^{+0.2}_{-0.3}$ 	& 2.5$^{+0.7}_{-0.6}$ \\
$kT_{\rm e}$ (keV) & 
13.3$^{+1.4}_{-1.5}$ 	& 13.6$^{+0.9}_{-1.7}$ 	& 15.0$^{+2.2}_{-2.5}$ 	& 8.2$^{+2.4}_{-1.7}$ 	& 5.0$^{+3.7}_{-1.7}$ \\
Energy$_{\rm Fe\emissiontype{XXV}}$ (keV) & 
6.33$\pm$0.03 			& 6.34$^{+0.03}_{-0.02}$	& 6.34$\pm$0.03 		&6.31$^{+0.04}_{-0.03}$	& 6.33 (fixed) \\
Intensity$_{\rm Fe\emissiontype{XXV}}$ ($\times$10$^{-5}$ photons s$^{-1}$ cm$^{-2}$) & 
0.9$^{+0.2}_{-0.3}$ 		& 1.8$^{+0.3}_{-0.4}$ 	&1.3$\pm$0.4 			& 2.9$\pm$0.8			& $<$1.7\\
Energy$_{\rm Fe\emissiontype{XXVI}}$ (keV) & 
6.60$^{+0.04}_{-0.05}$	& 6.62$^{+0.04}_{-0.05}$  & 6.63$^{+0.13}_{-0.08}$	& 6.60$^{+0.06}_{-0.08}$	& 6.61 (fixed)\\
Intensity$_{\rm Fe\emissiontype{XXVI}}$ ($\times$10$^{-5}$ photons s$^{-1}$ cm$^{-2}$) & 
0.6$\pm$0.3			& 0.9$^{+0.3}_{-0.4}$	& 0.4$^{+0.4}_{-0.3}$ 	&1.2$\pm$0.7 			& $<$1.5\\
$\chi^2$/d.o.f. &538/490 	& 1050/970 			& 468/450				&186/211 				& 197/146 \\
\hline \\
\end{tabular}
\end{center}
\end{table*}

In order to examine the spatial variation, spectra are extracted from 5 annular regions of 
0$''$--100$''$, 100$''$--200$''$, 200$''$--300$''$, 300$''$--500$''$, and 500$''$--700$''$. 
Point sources with $>$2$\times$10$^{-14}$ erg s$^{-1}$ cm$^{-2}$ in the 0.4--2.3 keV energy band (reported in \cite{Barriere2015}) are excluded. 
The source regions are displayed in figure 1. 

The spectra are fitted with a model consisting of the non-X-ray background (NXB), GDXE, CXB, and emission from IGR J17448$-$3232. 
According to \citet{Snowden2008}, the NXB is modeled as a PL function representing the soft proton contamination 
(a time variable flare component and a quiescent continuum component)
plus Gaussians representing instrumental fluorescence lines. 
The normalization and index of the PL function and normalizations of the Gaussians were set to be free. 
Since the intensities of the LP, HP, and RC components of the GDXE depend on the Galactic latitude, 
the intensities are adjusted to those at the position of IGR J17448$-$3232 according to \citet{Yamauchi2016}. 
Taking account of the scale heights of the 6.4 and 6.7 keV lines, 
we scaled down the intensities of RC and thermal components (LP and HP) by 0.80 and 0.89, respectively. 
On the other hand, since the spatial distribution of the FE component of the GDXE has not been well studied, 
there is large ambiguity.
Thus, we used data above 1.2 keV where the contribution of the FE component is low.  
The spectral parameters of the CXB were fixed to the values in \citet{Kushino2002}. 
The contribution of the GDXE and the CXB was in the range from $\sim$3 \% (0$''$--300$''$ region) to $\sim$27 \% (500$''$--700$''$ region) of the total count rates. 


The spectrum of IGR J17448$-$3232 clearly exhibits Fe\emissiontype{XXV} He$\alpha$ and Fe\emissiontype{XXVI} Ly$\alpha$ lines, as is shown in figure 3. 
In order to estimate an ionization temperature, 
we simultaneously fitted the MOS and pn spectra in the 1.2--11 keV band 
with a model of thermal bremsstrahlung ({\tt zbremss} in XSPEC) plus Gaussian lines 
with fixing the redshift of {\tt zbremss} to 0.055 \citep{Barriere2015}. 
Free parameters are normalization and electron temperature of {\tt zbremss}, a center energy and an intensity of Gaussian lines, and a column density. 
The best-fit parameters are listed in table 2 and the best-fit model is plotted in figure 3. 
Some residuals due to the instrumental line were found at $\sim$8 keV for the pn, which does not affect the results. 
The other instrumental lines of the MOS and the pn were well reproduced.  

Figure 4 displays a correlation plot between the electron temperature and 
the intensity ratio of Fe\emissiontype{XXVI} Ly$\alpha$ line/Fe\emissiontype{XXV} He$\alpha$ lines.
The ratios in the inner regions ($r<$300$''$) are 0.3--0.7, which corresponds to that of $\sim$8--10 keV plasma in the CIE state, 
while the temperature estimated from the continuum shape is 13--15 keV.
On the other hand, the ratios in the outer region ($r>$300$''$) are consistent with that in the CIE state. 
In order to check influence of the background estimation on the intensity ratio of the lines and the electron temperature, 
we varied the intensity of the sky background and the detector background by $\pm$10 \% and then obtained the consistent results within the statistical errors. 

\begin{figure}
  \begin{center}
\includegraphics[width=8cm]{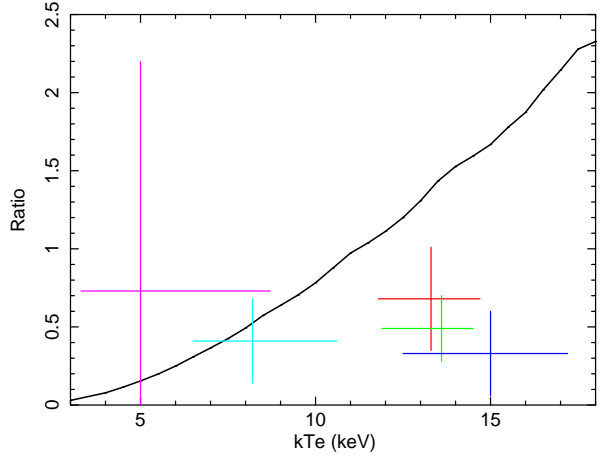} 
  \end{center}
  \caption{
A correlation plot between electron temperature and 
a ratio of Fe\emissiontype{XXVI} Ly$\alpha$ line/Fe\emissiontype{XXV} He$\alpha$ line: 
red, green, blue, cyan, and magenta colors show data points of 0$''$--100$''$, 100$''$--200$''$, 200$''$--300$''$, 
300$''$--500$''$, and 500$''$--700$''$, respectively (see table 2). 
The black line indicates the case of  the CIE plasma. 
}\label{fig:img}
\end{figure}



As shown in figure 4, the spectra within 300$''$ exhibit the NEI feature.
We fitted the spectrum extracted from a circle with a radius of 300$''$ with a single temperature CIE model and found that 
the CIE model does not represent the ratio of Fe\emissiontype{XXVI} Ly$\alpha$ line/Fe\emissiontype{XXV} He$\alpha$ line well. 
The spectrum and the best-fit model is shown in figure 5a, while the best-fit parameters are listed in table 3 ($\chi^2$/d.o.f.=1125/1120=1.004). 
Thus, we next examined an NEI plasma model ({\tt vrnei} in XSPEC) for the spectrum within 300$''$. 
The initial electron temperature is assumed to be 5 keV that is in the outermost region (see table 2). 
The NEI model well represented the intensity ratio of Fe\emissiontype{XXVI} Ly$\alpha$/Fe\emissiontype{XXV} He$\alpha$ lines.
The best-fit parameters are listed in table 3 ($\chi^2$/d.o.f.=1083/1119=0.968) and the best-fit model is plotted in figure 5b. 
The $n_{\rm e}t$ value is estimated to be (2.5$^{+1.4}_{-1.0}$)$\times$10$^{11}$ cm$^{-3}$ s. 

%
\begin{table*}[t]
\caption{The best-fit parameters for the $r<$300$''$ spectrum.}
\begin{center}
\begin{tabular}{lcccc} \hline  
Parameter & \multicolumn{4}{c}{Value}\\ 
\hline 
Model &     CIE & NEI & 2 CIE & CIE$+$power law \\
\hline 
$N_{\rm H}$ ($\times10^{22}$ cm$^{-2}$) 		&2.06$^{+0.08}_{-0.05}$ 	& 1.96$\pm$0.05 		& 1.97$^{+0.04}_{-0.05}$ 		& 1.97$^{+0.06}_{-0.04}$ \\
Normalization$^{\ast}$					&3.13$\pm$0.06 		& 3.17$\pm$0.06 		& --- 						& --- \\
$kT_{\rm e}$ (keV) 						&11.8$^{+0.8}_{-1.0}$ 	& 14.1$^{+1.3}_{-1.1}$ 	& --- 						& --- \\
$kT_{\rm init}$ (keV) 					& --- 					& 5.0 (fixed) 			& --- 						& --- \\
$n_{\rm e}t$ ($\times$10$^{11}$ cm$^{-3}$ s) 	& --- 					& 2.5$^{+1.4}_{-1.0}$ 	& --- 						& --- \\
Normalization$_{\rm low}^{\ast}$			& --- 					& --- 					& 1.9$\pm$0.3				& 2.5$^{+0.3}_{-0.6}$  \\
$kT_{\rm e, low}$ (keV) 					& --- 					& --- 					& 7.9$^{+1.7}_{-1.1}$		& 9.4$^{+0.9}_{-1.5}$  \\
Normalization$_{\rm high}^{\ast}$			& --- 					& --- 					& 1.6$^{+0.3}_{-0.6}$ 		& ---\\
$kT_{\rm e, high}$ (keV) 					& --- 					& --- 					& $>$34 					& --- \\
Redshift 								& 0.055 (fixed) 			& 0.055 (fixed) 			&  0.055 (fixed)				& 0.055 (fixed)  \\
Abundance$^{\dag}$ 					&0.33$\pm$0.04 		& 0.19$\pm$0.04		& 0.36$\pm$0.05 			& 0.33$^{+0.05}_{-0.04}$\\
Normalization$_{\rm PL}^{\ddag}$			& --- 					& --- 					&  --- 					& 1.0$^{+1.2}_{-0.6}$$\times$10$^{-3}$  \\
$\Gamma$							& --- 					& --- 					&   --- 					& 1.1$^{+0.3}_{-0.4}$ \\
$\chi^2$/d.o.f. 							&1125/1120 			& 1083/1119			& 1098/1118 				& 1102/1118\\
\hline \\
\end{tabular}
\end{center}
\vspace{-12pt}
$^{\ast}$ Defined as 
10$^{-12}$$\times$$\int n_{\rm H} n_{\rm e} dV$ / [4$\pi D_A^2(1+z)^2$] (cm$^{-5}$),
where $n_{\rm H}$, $n_{\rm e}$, and $D_A$ are hydrogen density (cm$^{-3}$), electron density (cm$^{-3}$), and
angular distance (cm), respectively. \\
$^{\dag}$ Relative to the solar value \citep{Anders1989}.\\
$^{\ddag}$ In units of photons s$^{-1}$ cm$^{-2}$ keV$^{-1}$ at 1 keV.
\end{table*}

\begin{figure*}
  \begin{center}
\includegraphics[width=8cm]{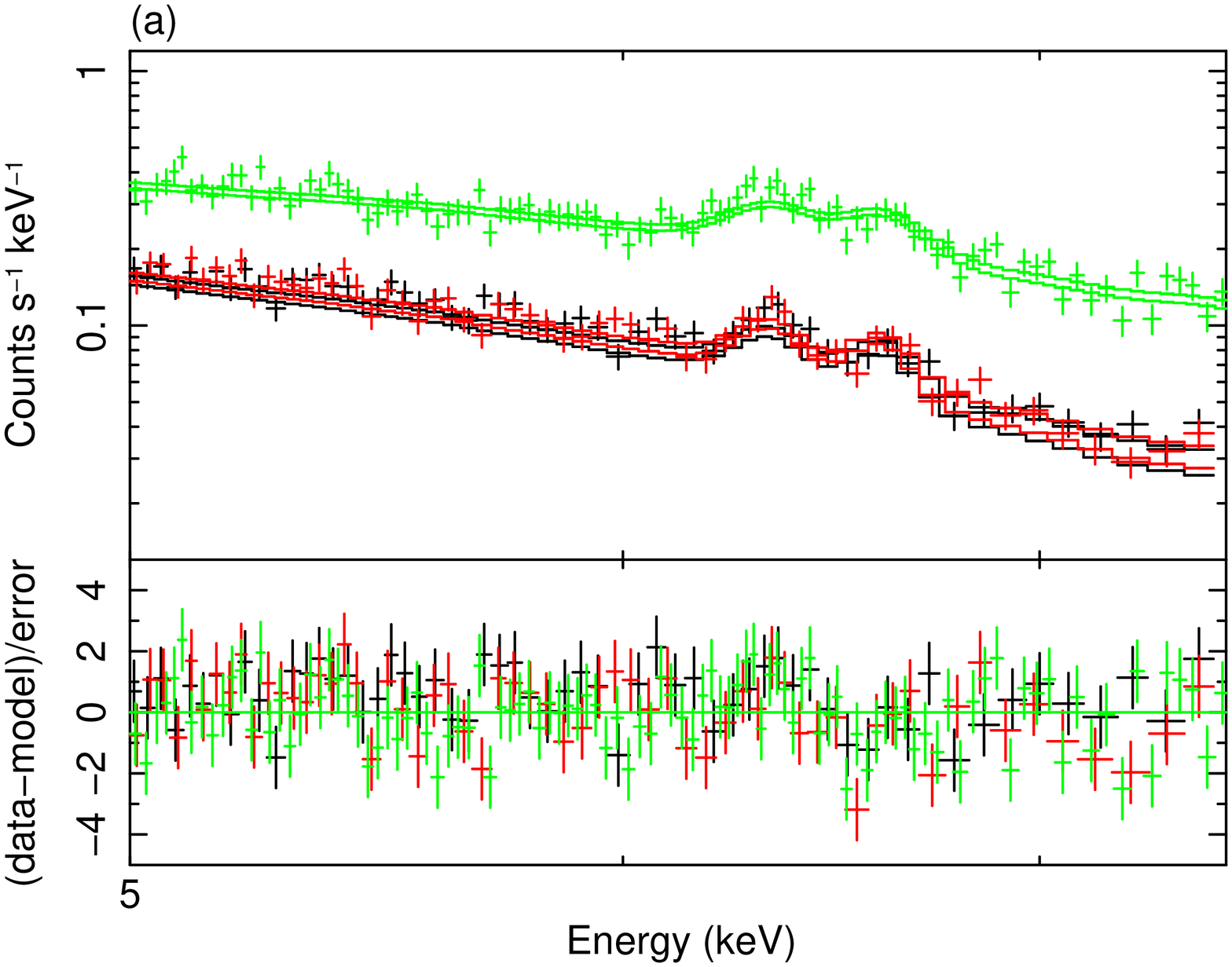} 
\includegraphics[width=8cm]{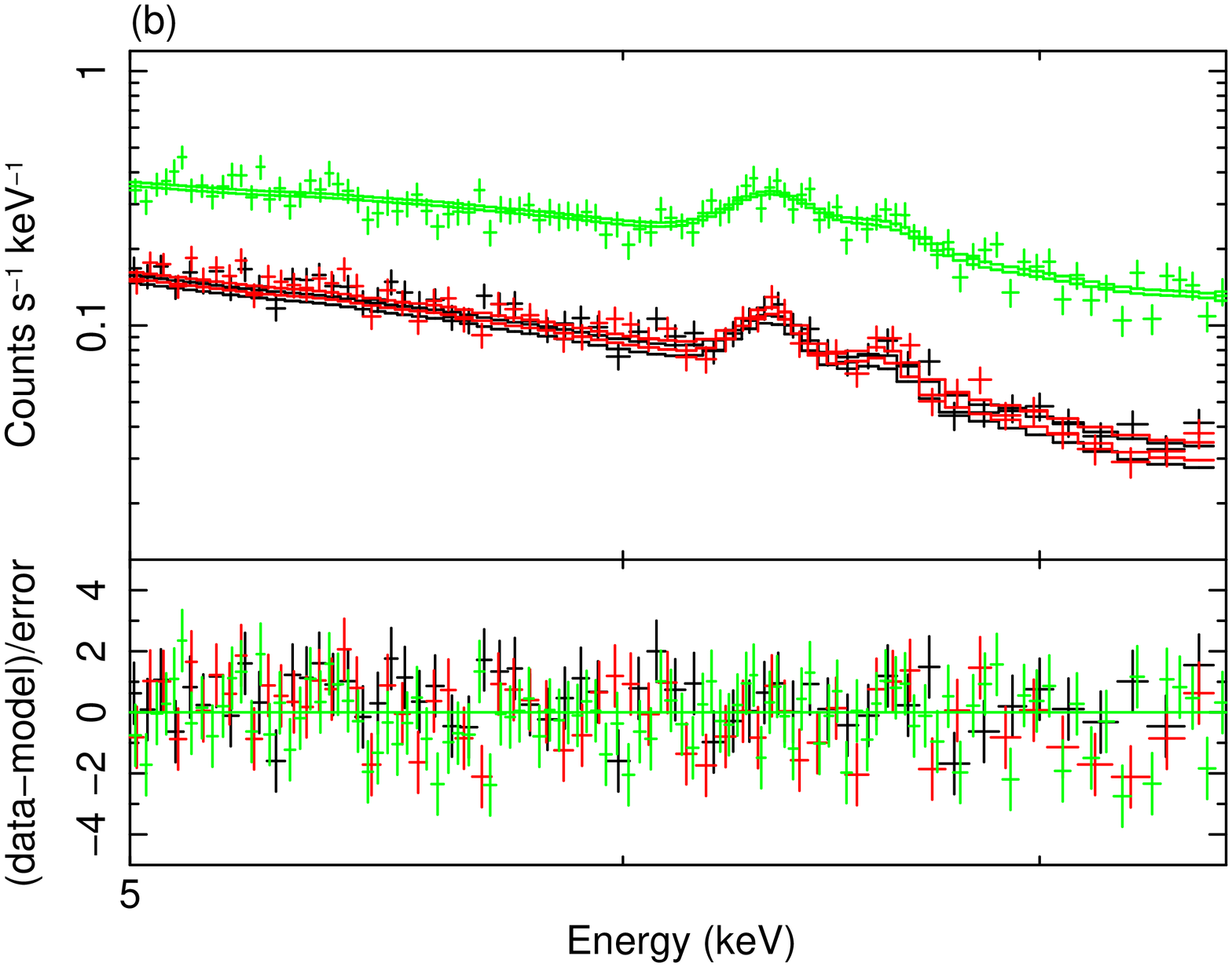} 
\includegraphics[width=8cm]{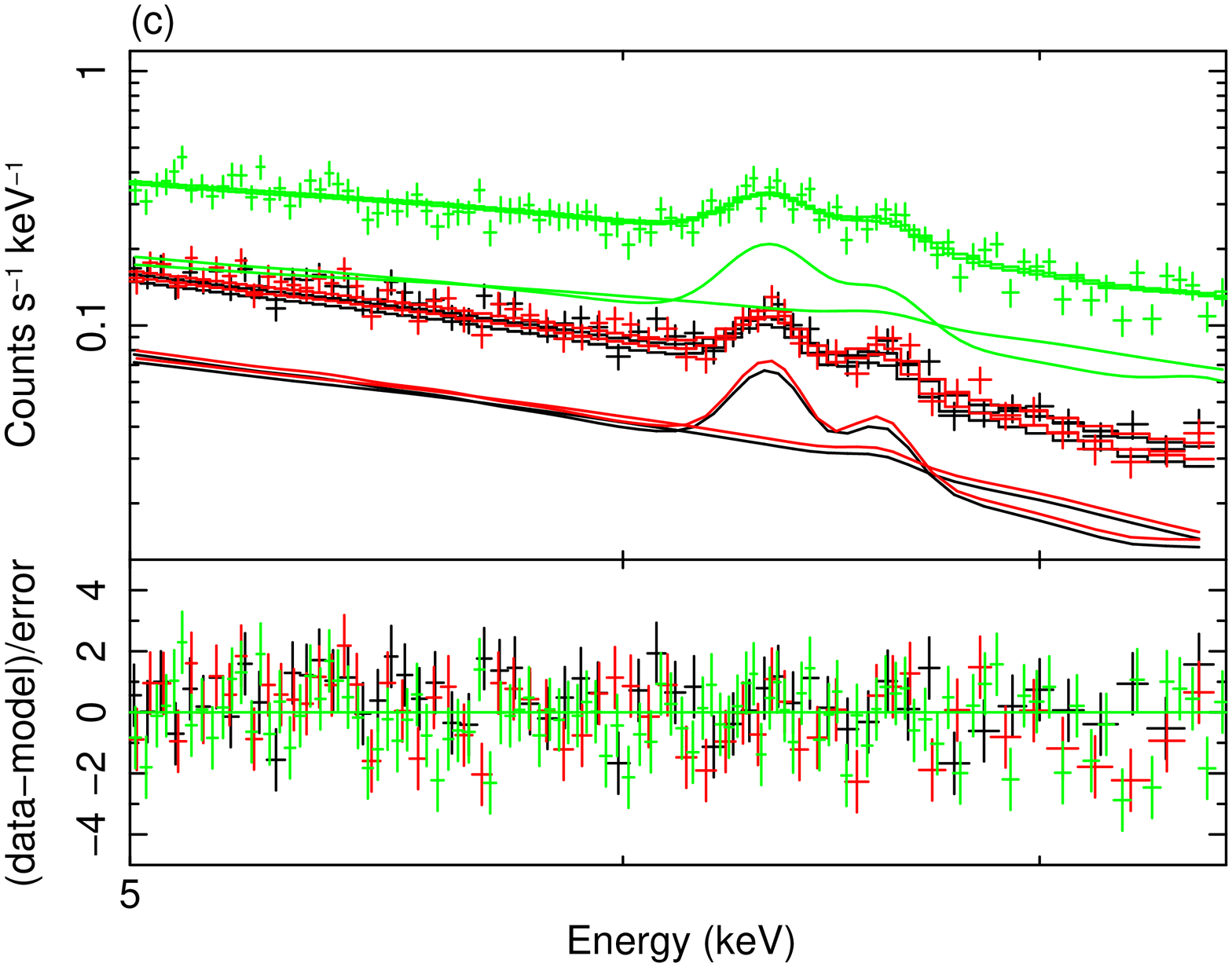} 
\includegraphics[width=8cm]{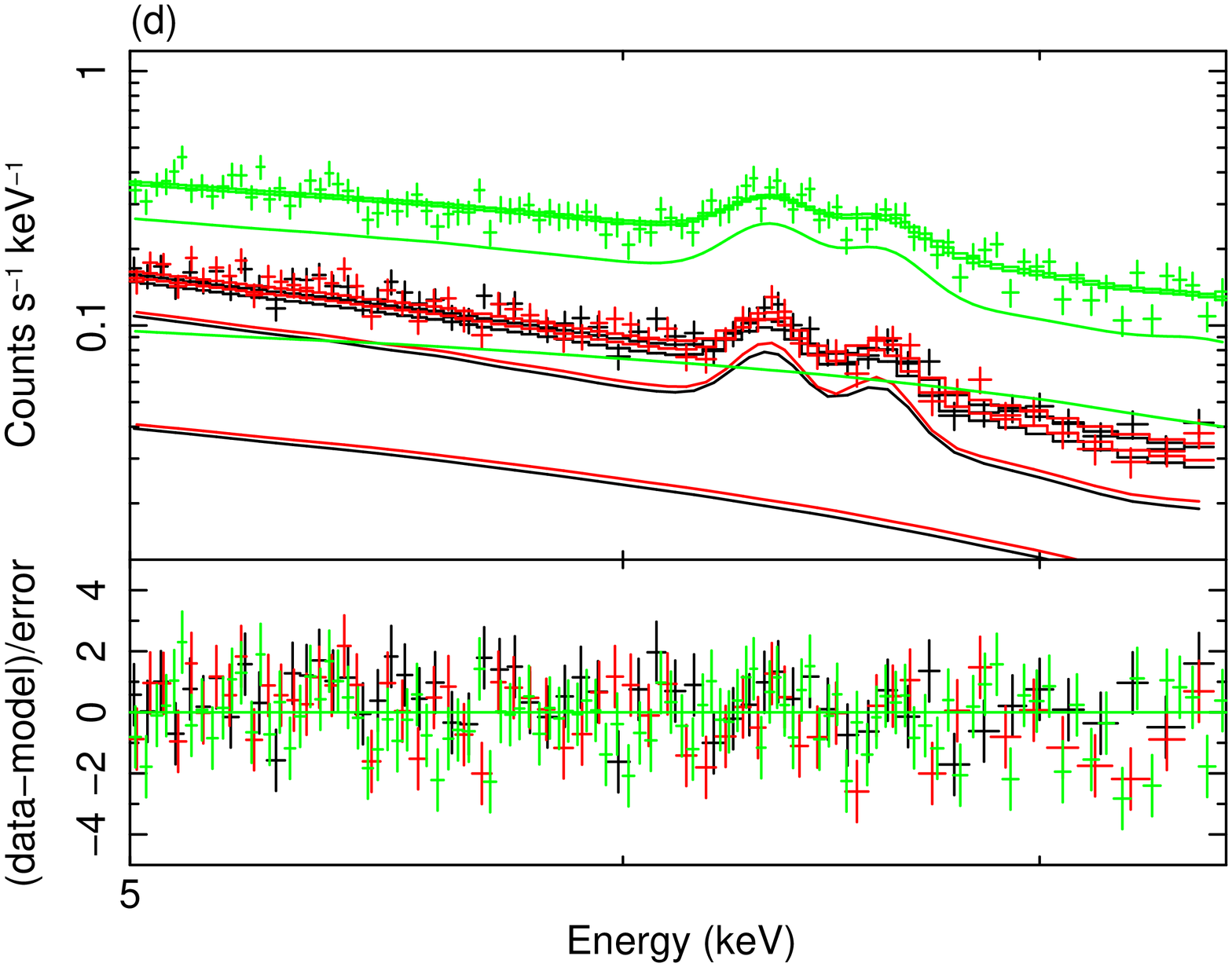} 
  \end{center}
  \caption{
X-ray spectra of IGR J17448$-$3232 extracted from a circle with a radius of of 300$''$ centered on the X-ray peak (upper panel) 
and residuals from the best-fit model (lower panel): (a) CIE, (b) NEI, (c) 2CIE, and (d) CIE+PL models. 
Although the model fitting was carried out for 1.2--11 keV band spectra, the 5--7.5 keV band data are displayed for the brevity. 
Black, red, and green colors show MOS 1, MOS 2, and pn, respectively.
Errors of the data points are at the 1 $\sigma$ level. 
}\label{fig:img}
\end{figure*}


Although the single-temperature NEI model represents the spectrum well, 
a multi-component model might be another possibility.
We also carried out a spectral fitting using a two-temperature CIE model (2CIE model). 
The abundances are linked between the low and the high temperature components.  
The model gave an acceptable fit ($\chi^2$/d.o.f.=1098/1118=0.982). The best-fit parameters are listed in table 3 and the best-fit model is plotted in 
figure 5c.
The obtained temperatures are 7.9$^{+1.7}_{-1.1}$ keV and $>$34 keV. 
In the case that the abundances are unlinked, we found that the abundance of the high temperature component cannot be constrained ($Z<0.6$ solar) 
and the improvement is low ($\Delta \chi^2$ of $\sim$2). 

We also applied a CIE$+$PL model and found that the model also gave an acceptable fit ($\chi^2$/d.o.f.=1102/1118=0.986).
The best-fit parameters are listed in table 3 and the best-fit model is plotted in figure 5d. 
The best-fit temperature and photon index are 9.4$^{+0.9}_{-1.5}$ keV and 1.1$^{+0.3}_{-0.4}$, respectively. 

We carried out the same model fit using the other abundance table in \citet{Lodders2003} and obtained consistent results within given statistical errors.


\section{Discussion}

IGR J17448$-$3232 roughly has a circular structure with no substructure (see figure 1). 
When a single-temperature plasma is assumed, 
the ICM exhibits a high electron temperature of 13--15 keV in the central region and a low electron temperature of 5--8 keV in the outer region and
has no evidence of a cool core that found in Ophiuchus cluster \citep{Fujita2008}. 
We found a sign that the intensity ratio of Fe\emissiontype{XXVI} Ly$\alpha$/Fe\emissiontype{XXV} He$\alpha$ lines in the central region is 
lower than that of the CIE plasma with a temperature of 13--15 keV.
The spectrum of $r<$300$''$ can be fitted with not only the single-temperature NEI model but also the two-component model (see table 3). 

When the 2CIE model is applied, the electron temperatures are estimated to be 7.9 keV and $>$34 keV (table 3). 
In the case of the CIE$+$PL model, the electron temperature of the thermal component and the photon index of the PL model are 9.4 keV and 1.1, respectively.  
The temperatures of 8--10 keV are consistent with those derived from the Fe\emissiontype{XXVI} Ly$\alpha$/Fe\emissiontype{XXV} He$\alpha$ line ratio 
(see figures 5c and 5d).

Using the spatial distribution derived from the isothermal $\beta$ model fit, 
the bolometric luminosities within a cluster radius, $R_{\rm 500}$ (a radius with a mean overdensity of 500 times the critical density of the Universe), 
of the 8 keV (2CIE) or 9 keV (CIE$+$PL) component are estimated to be 
$\sim$8$\times$10$^{44}$ erg s$^{-1}$ or $\sim$1$\times$10$^{45}$ erg s$^{-1}$, respectively, 
which are lower than those expected from the $L_{\rm X}$--$T$ relation (e.g., \cite{reichert2011}). 
As for the hard components, 
the electron temperature of the hard component from the 2CIE model fit is higher than typical values of the relaxed (e.g., \cite{reichert2011}) and merging galaxy clusters 
(e.g., 1E 0657$-$56: \cite{Markevitch2002}; RX J1347.5$-$1145: \cite{Ota2008}), while the photon index of 1.1 from the CIE$+$PL model fit is quite hard. 
In addition, the 2--10 keV band flux of the hard component corresponds to about a half (2CIE) or about one-fourth (CIE$+$PL) of the total flux. 
Such a high flux of the high-temperature or non-thermal component relative to the thermal one has not been reported so far. 
Furthermore, the NEI model gives a better fit than the 2CIE or CIE$+$PL model although all the models are statistically acceptable. 
Based on these results, we argue that the NEI plasma model in the central region is likely rather than the 2CIE or the CIE$+$PL model.

The NEI plasma in the central region would be produced by shock-heating via a merger in the past. 
Using the obtained plasma parameters, we estimate an elapsed time from the shock-heating. 
A volume emission measure of $r<$300$''$ is calculated to be $\sim$2$\times$10$^{67}$ cm$^{-3}$. 
Assuming a cylindrical structure with a radius of 300$''$ and a height of 2$R_{\rm 500}$, the electron density, $n_{\rm e}$, 
is estimated to be $\sim$10$^{-3}$ cm$^{-3}$. 
Hence, we can calculate an elapsed time to be $t_{\rm elapse}$$\sim$$n_{\rm e}t$/$n_{\rm e}$$\sim$8$\times$10$^{6}$ yr, which suggests 
the merging event would be occurred $\sim$8$\times$10$^{6}$ yr ago.
The crossing time of the cluster, $t_{\rm cross}$, by a sound velocity, $c_{\rm s}$, is estimated to be $t_{\rm cross}$=2$R_{\rm 500}$/$c_{\rm s}$$\sim$10$^9$ yr.
Since $t_{\rm cross}$ is much longer than $t_{\rm elapse}$, it is unlikely that the merger event was occurred on the celestial plane. 
IGR J17448$-$3232 has a circular structure without substructures and only the inner part has the NEI plasma with a high electron temperature. 
Based on these facts, we speculate that the merging event of IGR J17448$-$3232 is occurred along a line-of-sight direction. 
In order to check the scenario, further observational and theoretical studies are encouraged.

\section{Conclusion}

We analyzed the XMM-Newton data of the galaxy cluster IGR J17448$-$3232. 
The results are summarized as follows.

\begin{itemize}

\item The X-ray emission has a circular structure with no substructure. 

\item The ICM in the central region exhibits high electron temperature of 13--15 keV, while that in the outer region the temperature is 5--8 keV.

\item An ionization temperature is estimated from an intensity ratio of Fe\emissiontype{XXVI} Ly$\alpha$/Fe\emissiontype{XXV} He$\alpha$ lines. 
The value in the inner region is lower than 13 keV, while that in the outer region is consistent with those in the CIE state. 
These suggest that the ICM in the inner region is in the NEI state. 

\item The spectrum in the central region can be also represented by a two-component model:  
a two-temperature CIE model with temperatures of 7.9 keV and $>$34 keV 
or a CIE$+$power law model with a temperature of 9.4 keV and a photon index of 1.1. 
The high-temperature component or the power law component is quite hard. 

\item The obtained results support an idea that IGR J17448$-$3232 is a merging galaxy cluster. 
We speculate that the merging event of IGR J17448$-$3232 is occurred along the line-of-sight direction. 

\end{itemize}

\vspace{1pc}



\end{document}